\documentstyle[psfig]{l-aa}

\begin{document}

\thesaurus {06 (10.15.1, 10.15.2 (NGC 6604))}

\title{Spectroscopy and $BVI_{\rm C}$ photometry of the young open cluster 
       NGC~6604\thanks{Based on observations secured with the telescopes
       of the South African Astronomical Observatory (SAAO) and those of the
       Padova and Asiago Astronomical Observatories}}

\author{R. Barbon\inst{1}
\and   G. Carraro\inst{1}
\and   U. Munari\inst{2}
\and   T. Zwitter\inst{3}
\and   L. Tomasella\inst{2}   
       }
\offprints{R.Barbon (barbon@pd.astro.it)}

\institute {
Osservatorio Astrofisico del Dipartimento di Astronomia, 
Universit\`a di Padova, I-36012 Asiago (VI), Italy
\and
Osservatorio Astronomico di Padova, Sede di Asiago, 
I-36012 Asiago (VI), Italy
\and
University of Ljubljana, Department of Physics, Jadranska 19,
1000 Ljubljana, Slovenia
}
\date{Received date..............; accepted date................}

\maketitle

\begin{abstract}
$BVI_{\rm C}$ photometry and classification spectroscopy is presented 
for the young open cluster NGC~6604. Additional Echelle spectroscopy of the
brightest members is used to check the reddening against the interstellar
NaI and KI absorption lines, to measure the cluster radial velocity and to
derive the individual rotational velocities. We obtain 1.7 kpc as the
cluster distance, an age of $\sim 5\times 10^{6}$ years and a reddening
$E_{B-V}$= 1.02 ($\pm 0.01$ from three independent methods). Pre-ZAMS
objects are apparently not detected over the $\bigtriangleup m$ = 8.5 mag
explored range. The cluster radial velocity is in agreement with the Hron
(1987) model for the Galaxy disk rotation.
\end{abstract}
\maketitle

\section{Introduction}

NGC~6604 ($\alpha_{2000} =18^{h}~18^{m}.1$,
$\delta_{2000}=-12^{\circ}~14^{\prime}$, $l~=~18^{\circ}.3$,
$b~=~+1^{\circ}.7$) is a fairly compact open cluster belonging to class I3p
(Ruprecht 1966) and lies at the core of the HII region S54 (Georgelin et al.
1973). The fairly large extinction to the cluster ($A_{V} \sim
3$) is surely linked to the partnership with an HII region. 
The cluster contains several OB type stars (Stephenson and Sanduleak
1971), which suggest the cluster to be fairly young (Forbes and Dupuy 1978). 
Its distance is rather uncertain, with estimates ranging from 0.7 to 4.4 kpc
(Alter et al. 1970). Moffat and Vogt (1975, hereafter MV) presented
photoelectric photometry for a dozen stars in the cluster field, deriving a
1.6 kpc distance and a $E_{B-V}$=1.01 reddening. Forbes and DuPuy (1978,
hereafter FD) performed photoelectric and photographic photometry, which
yielded a 2.1 kpc distance, a $E_{B-V}$=0.96 reddening and a 4x10$^6$ years
age. They pointed out the possible presence of pre-ZAMS members which
demanded confirmation.

In this paper we report about our CCD $BVI_{\rm C}$ photometry of the
cluster central area, medium resolution spectroscopy of a field larger than
the cluster and high resolution Echelle spectroscopy of the brightest
cluster members.

\section {Photometry}

Observations of NGC 6604 have been carried out with the CCD camera mounted
on the 1.0 m telescope of the South Africa Astronomical Observatory (SAAO)
at Sutherland on July 2, 1992. The journal of observations is given in
Table~1 and a finding chart is presented in Figure~1. The surveyed area
corresponds to $2.1\times3.3$arcmin centered on the Of star HD 167971. The
reader is referred to Munari and Carraro (1995) for the observing technique
employed.  As usual, we observed several E-regions standard stars to
determine the extinction coefficients ($K_{B}=0.77\pm0.04$,
$K_{V}=0.52\pm0.01$ and $K_{I}=0.42\pm0.02$) and the color equations:

\begin{eqnarray}
B \ - \ b \ &=& \  -0.002(B-V) \ - 4.204 \\
V \ - \ v \ &=& \  -0.025(B-V) \ - 4.387 \\
I \ - \ i \ &=& \  -0.055(V-I) \ - 4.954 \\ 
B \ - \ V \ &=& \  +1.024(b-v) \ + 0.187 \\
V \ - \ I \ &=& \  +1.019(v-i) \ + 0.584 
\end{eqnarray}
\noindent
The resulting $BVI_C$ magnitudes are listed in Table~2 together with their
internal errors.  These ones average at $\sigma \sim$ 0.0015 mag and
increase only for stars fainter than $V = 14$ mag for which photon
statistics becomes the dominant source of noise.

Comparison between our CCD profile photometry (DAOPHOT) and the
photoelectric observations of MV yields for the seven stars in common:
\begin{eqnarray}
V \ - \ V_{MV} \ = \ 0.034  \ \ (\sigma=0.055) \\
(B-V)\ - \ (B-V)_{MV} \ = \ 0.033  \ \ (\sigma=0.020)
\end{eqnarray}

\begin{table}
\caption{Journal of observations. The last column is the FWHM of stellar 
images as measured on the CCD frames.}
\begin{flushleft}
\begin{tabular}{cccc} \hline
\multicolumn{1}{c}{Date} &
\multicolumn{1}{c}{Filter} &
\multicolumn{1}{c}{Exp. time} &
\multicolumn{1}{c}{Seeing} \\
 & &  (sec) & ($\prime\prime$) \\ \hline
July 9, 1992 & V &  30 &  2.1\\
& V & 180 &  2.2\\
& B & 300 &  2.4\\
& B &  60 &  2.3\\
& I &   3 &  2.1\\
& I &  30 &  1.9\\
\hline
\end{tabular}
\end{flushleft}
\end{table}

\begin{figure}
\centerline{\psfig{file=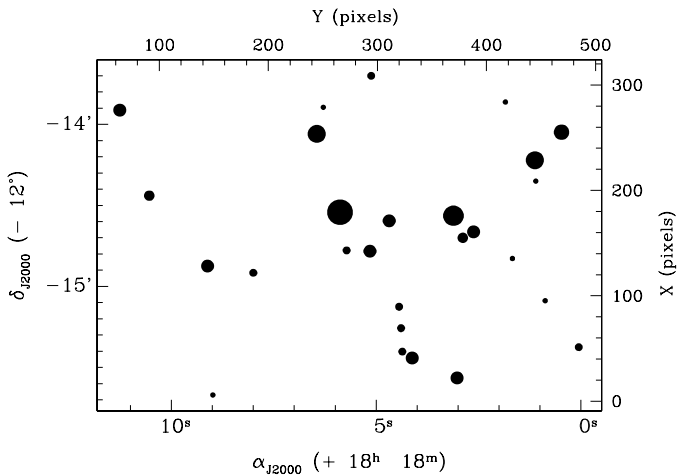,width=9cm}}
\caption[]{Identification map for the stars with $BVI_{\rm C}$ photometry in
Table~2. The imaged field covers $2.1\times3.3$ arcmin. North is up, East to
the left.}
\end{figure}

The comparison with the photoelectric observations of FD gives for the seven
stars in common:
\begin{eqnarray}
V \ - \ V_{FD} \ = \ 0.039 \ \ (\sigma=0.047) \\
(B-V)\ - \ (B-V)_{FD} \ = \ 0.092 \ \ (\sigma=0.028)
\end{eqnarray}
Our CCD profile photometry appears slightly fainter and redder than older
aperture photometry. It is our opinion that fainter and therefore generally
redder stars (quite abundant in a crowded field like that of NGC~6604) have
entered and contaminated the older aperture photometry, performed with fixed
diaphragms. The different standard stars used (Landolt' equatorial stars for
older photometric works, Cousin's E-regions for us) may add a little to the
differences in the photometry. The excellent agreement in the interstellar
reddening found below via three independent methods gives some support to such
arguments.

\begin{table*}
\caption{$BVI_{\rm C}$ photometry of NGC~6604. Columns gives our
identification number and those used by MV and FD, the X and Y positions on
the chip, the corresponding J2000.0 equatorial coordinates
(linked to the GSC~I system), and the BVI magnitudes and colours with
associated internal errors as provided by the DAOPHOT package.}
\begin{flushleft}
\begin{tabular}{rrrrrrrrrrrrr} \hline
\multicolumn{1}{c}{ID} &
\multicolumn{1}{c}{MV} &
\multicolumn{1}{c}{FD} &
\multicolumn{1}{c}{X}  & 
\multicolumn{1}{c}{Y}  &
\multicolumn{1}{c}{$\alpha_{J2000}$}  &
\multicolumn{1}{c}{$\delta_{J2000}$}  &
\multicolumn{1}{c}{$V$}& 
\multicolumn{1}{c}{$\sigma_V$} & 
\multicolumn{1}{c}{\sl B--V}   & 
\multicolumn{1}{c}{$\sigma_{B-V}$} & 
\multicolumn{1}{c}{\sl V--I} & 
\multicolumn{1}{c}{$\sigma_{V-I}$}  \\
\hline
  1&    &   D&     276.51&      63.89&  18 18  11.26 &--12 13  54.71 &12.101&      0.001&   0.763&      0.001&    1.025&    0.002 \\
  2&    &   M&     128.33&     144.03&  18 18  09.12 &--12 14  52.47 &12.201&      0.001&   0.800&      0.001&    1.065&    0.002 \\
  3&  11&    &     142.21&     293.03&  18 18  05.15 &--12 14  46.96 &12.993&      0.001&   0.840&      0.002&    1.271&    0.013 \\
  4&  10&    &     171.07&     310.88&  18 18  04.68 &--12 14  35.69 &12.996&      0.002&   0.807&      0.003&    1.108&    0.004 \\    
  5&    &  51&      69.04&     321.60&  18 18  04.39 &--12 15  15.48 &14.683&      0.005&   0.943&      0.009&    1.175&    0.004 \\
  6&    &    &      40.66&     331.84&  18 18  04.12 &--12 15  26.55 &12.873&      0.001&   0.839&      0.002&    1.138&    0.003 \\
  7&    &   J&      21.64&     373.07&  18 18  03.02 &--12 15  33.94 &12.227&      0.001&   0.807&      0.001&    1.094&    0.002 \\
  8&    &    &     154.74&     378.57&  18 18  02.88 &--12 14  42.02 &13.025&      0.002&   0.767&      0.003&    1.051&    0.004 \\
  9&    &    &     160.40&     388.27&  18 18  02.62 &--12 14  39.80 &12.411&      0.001&   0.751&      0.001&    1.052&    0.003 \\
 10&    &   2&     283.59&     417.69&  18 18  01.84 &--12 13  51.72 &15.894&      0.013&   1.020&      0.026&    2.586&    0.025 \\
 11&    &  72&      94.91&     453.76&  18 18  00.87 &--12 15  05.31 &15.987&      0.013&   1.026&      0.026&    1.394&    0.035 \\
 12&   9&   Z&     254.86&     469.18&  18 18  00.47 &--12 14  02.90 &11.658&      0.001&   0.772&      0.001&    1.057&    0.001 \\
 13&    &  52&       5.97&     148.61&  18 18  08.99 &--12 15  40.20 &15.187&      0.008&   0.943&      0.014&    1.236&    0.026 \\
 14&    &  49&     121.83&     185.94&  18 18  08.00 &--12 14  54.98 &14.971&      0.006&   1.176&      0.013&    1.505&    0.014 \\
 15&    &  31&     308.73&     294.42&  18 18  05.12 &--12 13  42.00 &14.478&      0.004&   0.989&      0.007&    1.185&    0.011 \\
 16&    &  71&     135.02&     424.02&  18 18  01.67 &--12 14  49.68 &15.282&      0.007&   0.909&      0.013&    1.200&    0.021 \\
 17&    &  76&      50.75&     484.57&  18 18  00.05 &--12 15  22.52 &14.259&      0.003&   0.872&      0.005&    1.239&    0.009 \\
 18&    &    &      46.77&     322.57&  18 18  04.36 &--12 15  24.17 &14.083&      0.003&   0.869&      0.006&    1.185&    0.009 \\
 19&    &    &     143.02&     271.89&  18 18  05.72 &--12 14  46.66 &14.265&      0.006&   0.892&      0.008&    1.271&    0.013 \\
 20&    &  50&      89.39&     319.79&  18 18  04.44 &--12 15  07.55 &14.790&      0.009&   0.963&      0.013&    1.240&    0.016 \\
 21&   2&   H&     253.65&     244.38&  18 18  06.45 &--12 14  03.52 &10.127&      0.001&   0.707&      0.001&    0.921&    0.001 \\
 22&    &    &     278.91&     250.70&  18 18  06.29 &--12 13  53.66 &15.454&      0.011&   1.037&      0.019&    1.238&    0.027 \\
 23&   3&    &     175.74&     369.74&  18 18  03.11 &--12 14  33.83 & 9.681&      0.001&   0.729&      0.001&    0.972&    0.001 \\
 24&   4&   Y&     228.36&     444.68&  18 18  01.12 &--12 14  13.25 &10.426&      0.001&   0.735&      0.001&    0.984&    0.001 \\
 25&    &    &     208.44&     445.42&  18 18  01.10 &--12 14  21.02 &15.112&      0.010&   0.853&      0.016&    1.233&    0.019 \\
 26&   1&   X&     179.19&     265.92&  18 18  05.88 &--12 14  32.55 & 7.524&      0.001&   0.812&      0.001&    1.116&    0.001 \\
 27&    &    &     195.26&      90.86&  18 18  10.54 &--12 14  26.39 &13.716&      0.017&        &           &    2.890&    0.017 \\
\hline 
\end{tabular}
\end{flushleft}
\end{table*}

\section{Low resolution spectroscopy}

Slitless spectroscopic observations of a circular area of $\sim 13$ arcmin
diameter centered on the cluster have been obtained with the AFOSC
spectrograph at the Asiago 1.8-m telescope on July 29 1998 between 20:00 and
22:30 UT (see Figure~2).  We used grism No.\ 7 (100 \AA / mm), resulting in
a dispersion of $\sim 2~$\AA/pix and a resolution (seeing/guide dominated)
of $\sim 7~$\AA. The night was photometric with a seeing better than 1.5
arcsec. Altogether 10 spectral frames bracketed by 17 V-band images were
obtained at different orientations (0$^\circ$, 45$^\circ$ and 90$^\circ$
position angles) and with different placements on the spectrograph focal
plane . This way the spectral overlap of stars near the clusters center can
be disentangled. All spectral frames have been exposed for 60 sec, except
for a 300 sec one.

The observations were reduced using IRAF 2.11 running on a
laptop PC under Linux operating system. One dimensional spectral tracings
were wavelength calibrated using the positions of the stars in the adjacent
$V$-band images as a reference. Once the spectral lines have been identified,
their centroids have been used for refinement of wavelength calibration.
Spectral information could be extracted for 32 stars, most of
them observed in multiple exposures. The wavelength calibration was
generally good to 2 \AA. The covered wavelength range for most stars was
$\lambda\lambda 3700 - 6500$\AA.  The flux calibration was obtained using
standard stars observed during the same night. Inter comparison of tracings
of the same star observed at different placements in the field and at
different spectrograph orientations shows that the relative flux calibration
(colours) are accurate to within 5\%\ redwards of 4500 \AA. The resolution
and the sometime low S/N ratio of the spectra has prevented in most cases
the determination of the luminosity class.
Our spectral types, $V$ magnitudes (from Table~2 when appropriate, or from 
the AFOSC $V$ band frames in the other cases), {\sl B--V} colors (from
Table~2 when appropriate, and in the other cases from convolution with the
appropriate photometric band profile on the flux calibrated spectra)
and other derived data are listed in Table~3, where comparison with 
the scarce spectral classification from literature is also provided
(Stephenson and Sanduleak 1971, Fitzgerald et al. 1979).

\section{Echelle Spectroscopy}

High resolution spectral observations of the four brightest stars in
NGC~6604 have been carried out on Oct 10, 1998 with the Echelle spectrograph
mounted at the Cassegrain focus of the 1.82 m telescope operated by
Osservatorio Astronomico di Padova on top of Mt.Ekar in Asiago (Italy). The
detector has been a THX31156 Thomson CCD 1024$\times$1024 pixels, 19$\mu$m
each and the slit was set to 2.2 arcsec. The whole $\lambda$ range
4550--8750 \AA\ was recorded on each frame, with a average dispersion of
0.173 \AA/pix at NaI~D.  Stars \#21 and \#32 have been observed with a
2$\times$2 binning. The spectra have been extracted and calibrated in a
standard fashion with IRAF running under Linux on a desk-top Pentium. The
spectra are part of a long term investigation of the kinematics of young
open clusters conducted in Asiago (see Munari and Tomasella 1999 for
details).

\begin{figure*}
\centerline{\psfig{file=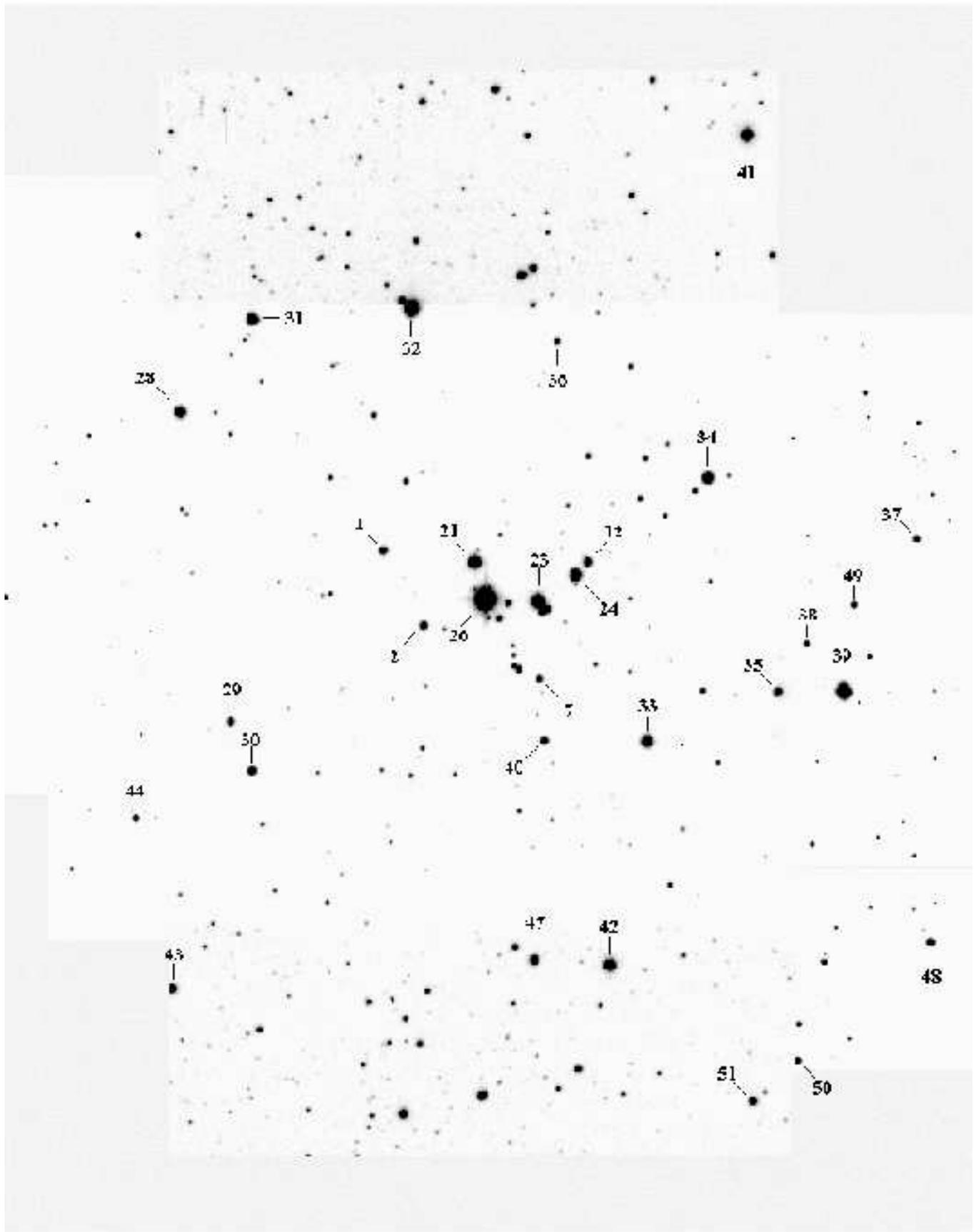,width=18cm}}
\caption[]{Identification map for low resolution slitless spectroscopy. North
is at the top, East to the left. The identification map for the $BVI_{\rm
C}$ photometry in Figure~1 covers the central part of this collage of $V$
band frames obtained in parallel with the slitless spectroscopic
observations.}
\end{figure*}

The profiles of the Na~I~D (5890, 5896 \AA) and KI (7699~\AA) interstellar
lines are presented in Figure~4 and their equivalent widths are given in
Table~4 together with the $E_{B-V}$ derived from the Munari and Zwitter
(1997) calibration.

The radial velocities of the same four cluster members are given in
Table~5, together with their rotational velocities 

\begin{figure}
\centerline{\psfig{file=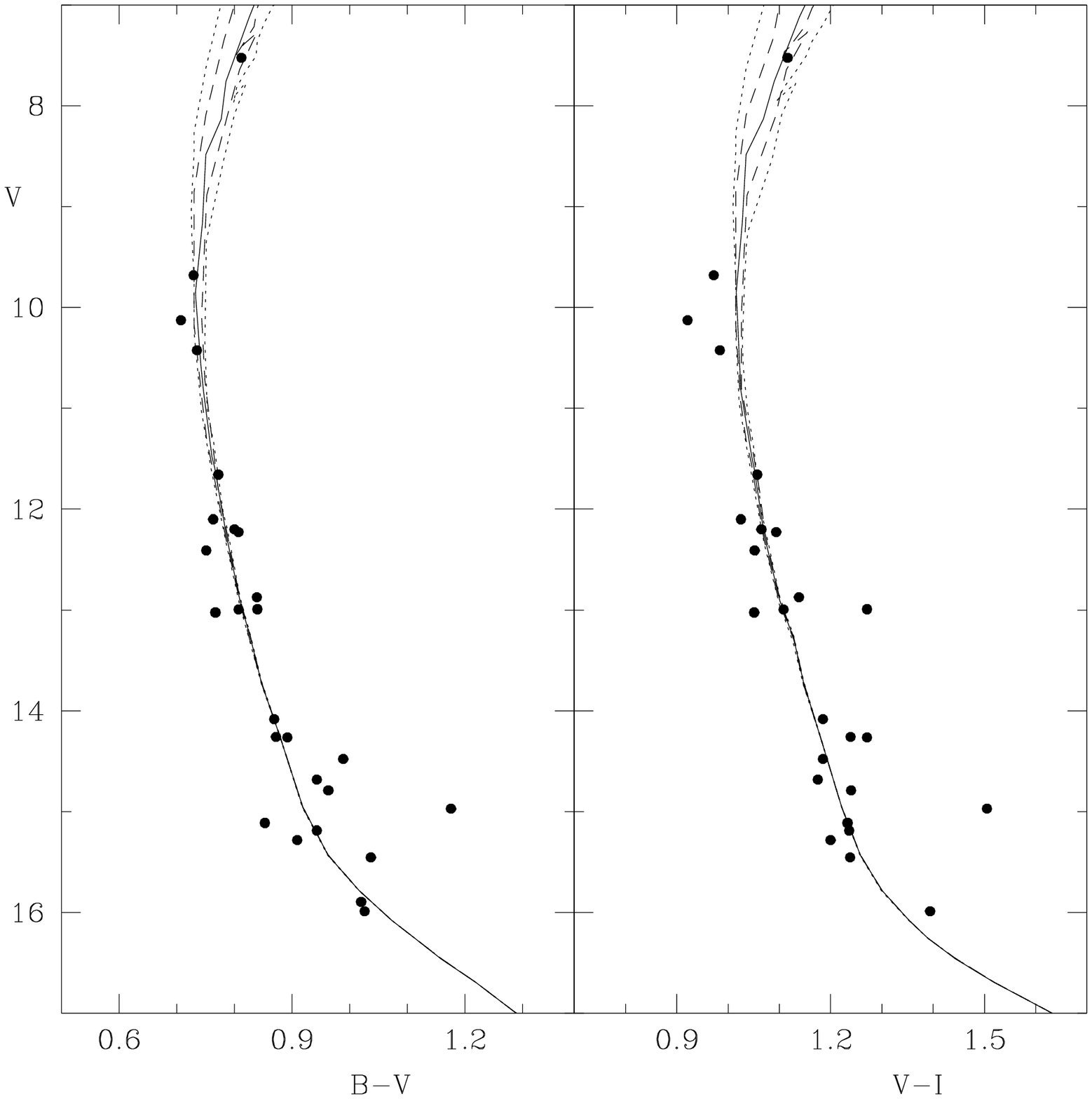,width=9cm}}
\caption[]{Colour-magnitude diagrams for the stars in Table~2. Solar 
abundance isochrones for 3,4,5,6 and 7$\times 10^{6} yr$ are overimposed.}
\end{figure}

\begin{table}
\caption{Our and literature spectroscopic classifications. 
Stars with identification numbers greater than 27 lie outside the
area covered by Figure~1 and are identified in Figure~3.}
\begin{flushleft}
\begin{tabular}{rllllll} 
\hline
\multicolumn{1}{c}{star}          & 
\multicolumn{1}{c}{$V$}           & 
\multicolumn{1}{c}{\sl B--V}      & 
\multicolumn{2}{c}{spectral type} & 
\multicolumn{1}{c}{$E_{B-V}$}     & 
\multicolumn{1}{c}{\sl V--M$_{V}$}\\
\cline{4-5}
&  &   & ours    & liter. &  &  \\
\hline
 1 & 12.101 & 0.763 & early B &      &      &        \\ 
 2 & 12.201 & 0.800 & B2V     &      & 1.04 & 14.65  \\ 
 7 & 12.227 & 0.807 & mid B   &      &      &        \\ 
12 & 11.658 & 0.772 & late B  &      &      &        \\ 
21 & 10.127 & 0.707 & B0      &  B0V & 1.01 & 14.13  \\ 
23 & ~9.681 & 0.729 & O9.5    & B1IV & 1.04 & 13.93  \\ 
24 & 10.426 & 0.735 & B1V     &      & 1.00 & 13.63  \\ 
26 & ~7.524 & 0.812 & O7      & O8f  & 1.10 & 14.32  \\ 
28 & 10.73  & 0.7   & B3      &      &      &        \\ 
29 & 12.27  & 0.7   & late B  &      &      &        \\ 
30 & 12.12  & 1.0   & G4      &      &      &        \\ 
31 & 11.06  & 1.0   & M4III   &      &      &        \\ 
32 & ~9.26  & 0.7   & O9.5    & B1Ib &      &        \\ 
33 & 10.43  & 0.9   & B4:     & OB   &      &        \\ 
34 & 10.82  & 0.9   & B0.5    & OB   &      &        \\ 
35 & 12.06  & 1.1   & early B &      &      &        \\ 
36 & 12.63  & 0.8   & early F &      &      &        \\ 
37 & 12.72  & 0.9   & K5:     &      &      &        \\ 
38 & 12.77  & 1.6   & late K  &      &      &        \\ 
39 & ~8.96  & 1.7   & early G &      &      &        \\ 
40 & 12.41  &       & K3      &      &      &        \\ 
41 & 10.65  &       & late B  &      &      &        \\ 
42 & 10.90  &       & early B &      &      &        \\ 
43 & 12.60  &       & K3      &      &      &        \\ 
44 & 12.66  &       & A       &      &      &        \\ 
45 & 12.83  &       & early F &      &      &        \\ 
46 & 12.65  &       & late A  &      &      &        \\ 
47 & 12.04  &       & A       &      &      &        \\ 
48 & 12.24  &       & K4      &      &      &        \\ 
49 & 12.94  &       & A       &      &      &        \\ 
50 & 12.22  &       & early G &      &      &        \\ 
51 & 12.90  &       & F:      &      &      &        \\ 
\hline
\end{tabular}
\end{flushleft}
\end{table}

\begin{table}
\caption{Equivalent width and corresponding $E_{B-V}$ (from the Munari \&
Zwitter 1997 calibration) of the interstellar NaI and KI lines in the
Asiago Echelle+CCD spectra of some NGC~6604 stars.}
\begin{flushleft}
\begin{tabular}{crccccc} 
\hline
star &  \multicolumn{1}{c}{$V$}   & \multicolumn{2}{c}{NaI (5890,5896 \AA)} 
                                  && KI (7699 \AA) & $E_{B-V}$ \\
\cline{3-4} \cline{6-6}
     &                            & EW (\AA)                     
                                  & EW (\AA) && EW (\AA)      &          \\
\hline
21 & 10.13 & 0.67 & 0.64 && 0.20      &1.0        \\ 
23 &  9.68 & 0.66 & 0.64 && 0.25      &1.0        \\ 
26 &  7.52 & 0.72 & 0.67 && 0.25      &1.1        \\ 
32 &  9.26 & 0.65 & 0.63 && 0.19      &0.9        \\ 
\hline
\end{tabular}
\end{flushleft}
\end{table}

\begin{table}
\caption{Radial and rotational velocity for some NGC~6604 stars derived from
Asiago Echelle+CCD spectra. {\sl EW, HIW} = equivalent width and half
intensity width of He~I $\lambda$ 5876 \AA; $V_{rot}\sin i$ = projected
rotational velocities derived from the HIW of the He~I $\lambda$ 5876 \AA\
line and the Munari \& Tomasella (1999) calibration.}
\begin{tabular}{ccccccr}
\hline
star & JD$_{\odot}$&\multicolumn{1}{c}{RV$_{\odot}$}&$\sigma$&\multicolumn{1}{c}{EW} &
\multicolumn{1}{c}{HIW} & \multicolumn{1}{c}{$V_{rot}\sin i$} \\  
&&(km/s)&(km/s)&(\AA)&(\AA)&(km/s)\\
\hline
&&&&&&\\
21& 2451097.277 & +21 & 2 & 1.35 & 8.81 & 354 $\pm$ 6 \\
23& 2451097.253 & +25 & 6 & 0.67 & 2.80 &  90 $\pm$ 5 \\
26& 2451097.234 & +15 & 3 & 0.64 & 3.10 & 104 $\pm$ 5 \\
32& 2451097.291 & +21 & 2 & 1.41 & 3.36 & 115 $\pm$ 6 \\
&&&&&&\\
\hline
\end{tabular}
\end{table}

\begin{figure}
\centerline{\psfig{file=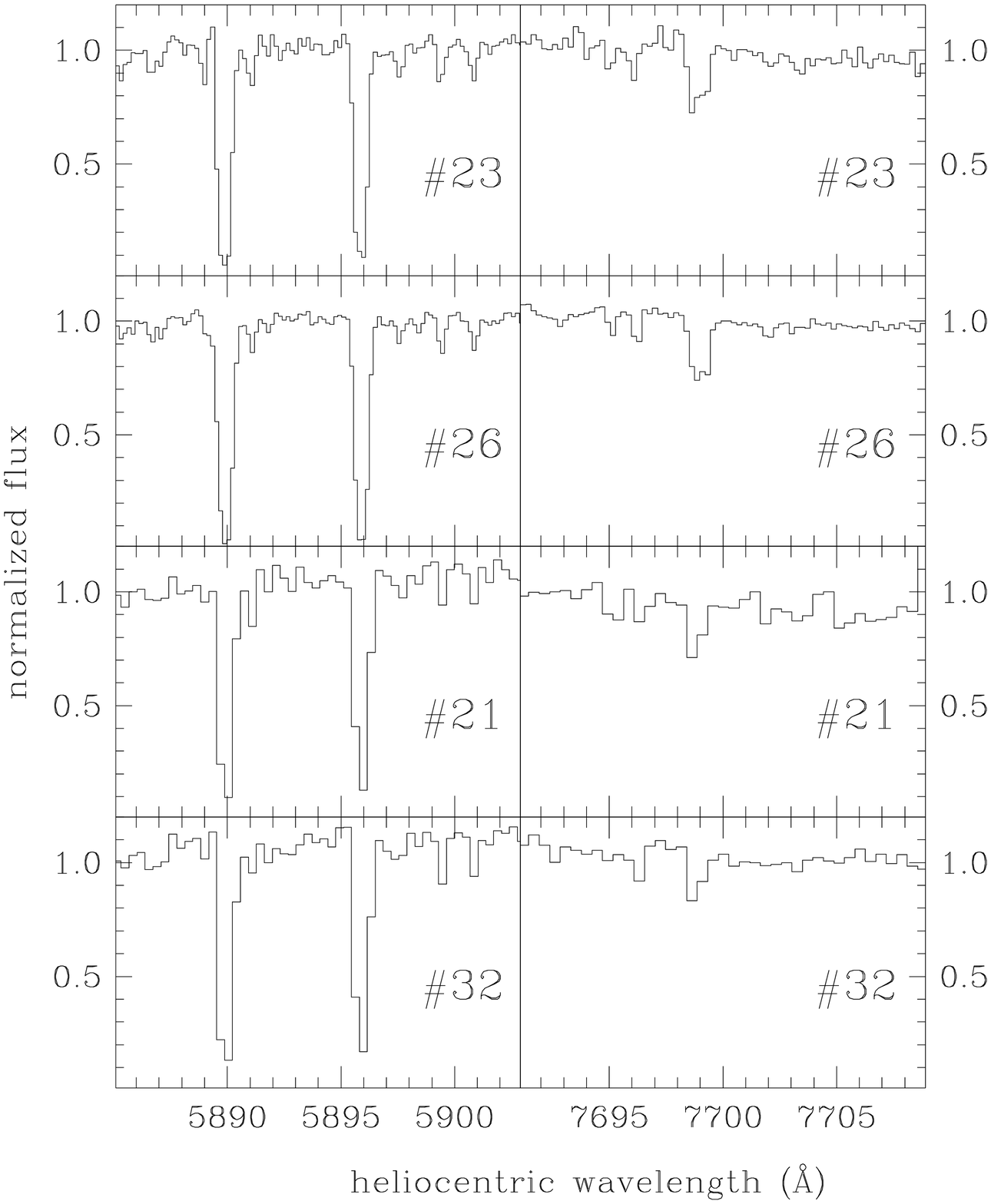,width=9cm}}
\caption[]{Interstellar NaI (5889.951, 5895.924 \AA) and KI (7698.979 \AA) 
lines from Asiago Echelle+CCD spectra of four NGC~6604 stars.}
\end{figure}

\section{Results}

The {\sl V,(B--V)} and {\sl V,(V--I)} diagrams for the stars in
Table~2 are shown in Figure~3. Stars \# 10 and 27, which have 
very red {\sl V--I} colors laying well outside the diagram borders
are not plotted. The situation for these two stars looks confuse
(and need additional observations to be settled), because 
\# 10 has a {\sl B--V} color that places it exactly on the ZAMS,
and \# 27 (which is too faint for a reliable measure on the $B$
frames) appears too bright in $V$ to be a pre-ZAMS cluster member.

We started with a fit of the observed $V,(B-V)$ diagram to the Zero Age Main
Sequence (Schmidt-Kaler 1982) obtaining an apparent distance modulus $m-M =
14.28$ and a mean reddening $E_{B-V}$ = 1.02. Then, the age was determined
through trial fits to the observed main sequence with theoretical isochrones
of the Padova group ( Bertelli et al. 1994) characterized by a standard
$[$He/H$]$ ratio and solar metallicity and scaled to the distance modulus
and reddening found before. The best fit, shown in Figure~3, is obtained for
the isochrone corresponding to the age of 5$\times 10^{6}$~years, slightly
older than the value reported by FD.  The resulting age is sensitive to the
position in the color-magnitude diagram of star \#26 (the brightest one) and
to a lesser extent of \#23. Both stars lie very close to the cluster center,
their spectro-photometric parallax and reddening agree with the cluster
distance and extinction, and furthermore the wide field spectroscopy of
Figure~2 and Table~3 shows a marked drop of O-B stars outside the region
covered by Figure~1. Thus, it is quite safe to assume both stars as
validated members of the cluster. The radial velocity of \#26 is $\sim 2.5
\sigma$ off the mean of the other three cluster members in Table~5, which
could suggest a binary nature. The contribution to the \#26 brightness by
the possible companion cannot be directly determined by the data at hand;
however it is worth noticing that ($a$) the companion is not severly
affecting the star colour, and ($b$) its spectral features does not show up
in our Echelle spectra. Thus the position of star \#26 in Figure~3 should
not be affected by a possible companion by more than a few tenths of a
magnitude. Dimming star \#26 by 0.25 mag would change by no more than 1
million year the age estimate, for which we can therefore assume a safe
5($\pm 2$)$\times 10^{6}$~years. It may be of interest to note that
Feinstein et al. (1986) have suggested that open clusters having stars with
Of characteristics like star \#~26 should not be older than $5\times10^{6}$
years,

The distance to NGC 6604 is d = 1.7~ kpc, for a R$_V$=3.1 standard reddening
law. Such a distance is 25\% smaller than found by FD, which is mainly based
on photographic photometry, but quite in agreement to the value derived by
MV and places the cluster at a galactocentric distance 6.9 kpc, on the outer
boundary of the Carina-Sagittarius arm. Adopting the ZAMS in the 
{\sl (B-V),(V-I)} as tabulated by Munari and Carraro (1996), a ratio 
\begin{equation}
\frac{E_{B-V}}{E_{V-I}} \ = \ 0.77
\end{equation}
is found for the reddening affecting NGC~6604, in agreement with the value
for the R$_V$=3.1 standard reddening law. The fitting to isochrones in the
{\sl V, V$-$I} plane is shown in Figure~3, with $V - M_V$ = 14.28 mag and
$E_{V-I}$=1.32 as scaling factors, in agreement with the values from the
{\sl V, B$-$V} diagram. In view of the larger uncertainities affecting the
transformations from the theoretical plane ({\sl log} L, {\sl log} T) to the
observational one when red photometric bands are involved, the resulting
isochrone fitting to the observational data seems satisfactory.

The spectroscopic data of Table~3 for the four stars with detailed spectral
classification (\#\ 2, 21, 23, 24) give a mean reddening of $E_{B-V}=
1.02\pm0.01$, the same determined from photometry (the star \# 26 has not
been considered because of its emission line nature). The mean reddening
from interstellar lines from Table~4 is $E_{B-V} = 1.00 \pm 0.04$. It seems
noteworthy that three independent methods converge within 0.01 mag to the
same $E_{B-V}= 1.02\pm0.01$ value for the reddening affecting NGC~6604.

The mean spectroscopic apparent modulus $V - M_{V}=14.13\pm0.17$ mag for the 
cluster members in Table~3 is in good agreement with the $V -
M_{V}=14.28$ derived from ZAMS fitting.

Finally, the data listed in Table~5 give a cluster heliocentric radial
velocity $RV_{\odot} = +20.5\pm 2.1$ km~sec$^{-1}$, in agreement with the
$RV_{\odot} =+19.0\pm 3.5$ km~sec$^{-1}$ of Liu et al. (1991).  The Hron's
(1987) rotation curve gives a heliocentric radial velocity of +8.2$\pm 2.5$
km~sec$^{-1}$ at the galactic location of NGC 6604. Bearing in mind that the
effect of the galactic rotation, as seen from the Sun, nearly vanishes
toward the Galaxy center direction (close to which NGC~6604 lies) the
resulting difference between model and observational velocities (12
km~sec$^{-1}$) is within the dispersion of the galactocentric radial
velocities for extreme Pop I objects (12.5 km~sec$^{-1}$, Binney and
Merrified 1998). Therefore the cluster distance and position, its radial
velocity and the Hron's model for the Galaxy disk rotation appear in good
mutual agreement.

\section {Conclusions}

Our determination for the cluster distance and reddening confirm earlier
studies. From the small dispersion around the cluster main sequence in
Figure~3 we may conclude that most of the stars in Figure~1 and Table~2 are
physical members of NGC~6604. From the spatial concentration of early
spectral types in Table~3, the diameter of NGC6604 is of the order of
$\sim$10 arcmin. The $E_{B-V}= 1.02\pm0.01$ reddening follows the standard
R$_V$=3.1 reddening law, with no evidence for a marked differential
reddening over this cluster.  The cluster age is estimated as
$5\times10^{6}$ years.  Our observations do not go faint enough to address
the reality of the pre-ZAMS objects suspected by FD. Extrapolating from the
turn-on tracks by Stauffer et al. (2000), pre-ZAMS objects in NGC 6604
should have $V\geq$ 16, corresponding to a ZAMS spectral type somewhat later
than A0. Thus, star \#14 (which is $\sim$2 mag off the ZAMS) looks too
bright both for a pre-ZAMS object and for an equal-mass binary, and
therefore it should be a field star.


\begin{thebibliography}{}

\bibitem{} Alter G., Ruprecht J., Vanysek J. 1970, Catalogue of Star Clusters and
           Associations, Budapest
\bibitem{} Bertelli G., Bressan A., Chiosi C., Fagotto F., Nasi E., 1994,
           A\&AS, 106, 275
\bibitem{} Binney J., Merrifield M., 1998, Galactic Astronomy, Princeton University Press
\bibitem{} Feinstein A., Vazques R. A., Benvenuto O. G., A\&A, 159, 223
\bibitem{} FitzGerald M. P., Luiken M., Maitzen H. M., Moffat A. F. J., 1979, A\&AS 37, 345
\bibitem{} Forbes D., DuPuy D., 1978, AJ, 83 266
\bibitem{} Georgelin Y. M., Georgelin Y.P., Roux S., 1973, A\&A 25, 337
\bibitem{} Hron J., 1987, A\&A 176, 34
\bibitem{} Liu T., Janes K. A., Bania T. M., 1991, AJ, 102, 1103
\bibitem{} Moffat A. F. J., Vogt N., 1975, A\&AS,  20,  155
\bibitem{} Munari U., Carraro G., 1995, MNRAS, 277, 1269
\bibitem{} Munari U., Carraro G., 1996, A\&A, 314, 108
\bibitem{} Munari U., Zwitter T., 1997, A\&A, 318, 269
\bibitem{} Munari U., Tomasella L., 1999, A\&A, 343, 806
\bibitem{} Schmidt-Kaler, Th. 1982 in Landolt-B\"{o}rnstein, Numerical Data
           and Functional Relationships in Science and Technology,
           Ed. K.Schaifer \& H.H. Voigt, New Series, Group IV, Vol.II(b) 
           (Pringer, Berlin), pag. 14
\bibitem{} Stauffer J.R., Jeffries R.D., Martin E.L., Terndrup D.M. 2000, 
           in "11th Cambridge Workshop on Cool Stars, Stellar Systems, and the Sun," ed. 
           R. J. Garcia Lopez, R. Rebolo, and M. R. Zapatero Osorio, in press. 
           See astro-ph/0001229
\bibitem{} Stephenson,C., B., Sanduleak N., 1971, Cat. of Luminous Stars in the 
           Southern Milky Way,  Warner\&Swasey Obs. Publ. Cleveland.
\bibitem{} Ruprecht J., 1966, Bull. Astron. Inst. Czech. 17, 34
\end{thebibliography}
\end{document}